\documentclass{article}
\usepackage{amsmath}
\usepackage{amsfonts}
\usepackage[margin=1in]{geometry}
\usepackage{graphicx}
\usepackage{multirow}
\usepackage{multicol}
\usepackage{booktabs}
\usepackage{xcolor}
\usepackage{indentfirst}
\usepackage{amsthm}
\usepackage{setspace}
\usepackage{bm}
\usepackage[notquote]{hanging}
\usepackage{amsthm}
\usepackage{rotating}
\usepackage{url}

\DeclareMathOperator{\logit}{logit}

\allowdisplaybreaks

\begin{document}

\title{Affiliation network model of HIV transmission in MSM}
\author{Jonathan Larson and Jukka-Pekka Onnela}
\date{\today}

\maketitle

\doublespacing

\begin{abstract}
Black men who have sex with men (MSM) in the U.S. are more likely
to be HIV-positive than White MSM.
Intentional and unintentional segregation of Black from non-Black MSM
in sex partner meeting places may perpetuate this disparity,
a fact that is ignored by current HIV risk indices,
which mainly focus on individual behaviors and not systemic factors.
This paper capitalizes on recent studies in which the venues
where MSM meet their sex partners are known.
Connecting individuals and venues leads to so-called affiliation networks;
we propose a model for how HIV might spread along these networks,
and we formulate a new risk index based on this model.
We test this new risk index on
an affiliation network of 466 African-American MSM
in Chicago, and in simulation.
The new risk index works well when there are two groups of people,
one with higher HIV prevalence than the other,
with limited overlap in where they meet their sex partners.

\end{abstract}

\section{Background}
\label{sec:intro}

In 2018, men who have sex with men (MSM) accounted for over 65\% of
new HIV infections in the U.S
\cite{cdc2020},
amounting to over 24,000 infections that year.
In almost every other country for which we have data,
MSM are several times more likely to be HIV-positive than the general population
\cite{beyrer2010}.
Also in 2018,
African-Americans accounted for over 40\% of new HIV infections in the U.S.,
%and at over 16,000 infections,
more than any other racial or ethnic group \cite{cdc2020}.
A 2012 meta-analysis \cite{millett2012}
found that the odds of HIV seropositivity in Black MSM
were three times that in non-Black MSM,
even though Black men were significantly less likely
to have unprotected anal intercourse (UAI) with a main male partner,
have a high number of male sex partners over their lifetime
or in the past year,
or report any lifetime or recent drug use.
The same meta-analysis
found that Black MSM were significantly more likely to use condoms,
undergo repeated HIV testing,
and use pre- or post-exposure prophylaxis
than non-Black MSM.
Clearly, individual risk behaviors are insufficient for explaining
the disparity in risk of HIV acquisition between Black and non-Black MSM.

This may explain the poor performance of measures designed
to estimate an individual's risk of contracting HIV,
which are largely based on individual behavior. 
For example,
Jones et al. \cite{jones2017} tested three previously published measures of risk
in a sample of Black and White MSM in Atlanta
and obtained AUCs ranging from $0.49$ to $0.63$ (among Black participants)
and from $0.60$ to $0.67$ (among White participants).
These measures were based on such characteristics as age,
total number of male partners,
number of episodes of condomless receptive anal intercourse,
amphetamine and popper use,
and diagnosis with gonorrhea, chlamydia, or syphilis.
Lancki, Almirol, Alon, McNulty, and Schneider \cite{lancki2018}
similarly examined
three risk indices in a sample of Black MSM in Chicago
and obtained AUCs ranging from $0.51$ to $0.57$.
One of these indices was the same as in Jones et al. \cite{jones2017};
one was based on similar factors such as condomless anal sex;
and one was based on similar factors as well as whether the individual
had exchanged sex for commodities or been incarcerated.

Evidence of the utility of a network-based approach is supported by 
Raymond and McFarland (\cite{raymond2009}, p. 630), who state,
``Black MSM were reported as the least preferred as sexual
partners,
believed at higher risk for HIV,
counted less often among friends,
were considered hardest to meet,
and perceived as less welcome at the common venues that cater to gay men
in San Francisco.''
Millett et al. \cite{millett2012} found that non-Black MSM
had less than a tenth of
the odds of reporting Black sex partners as Black MSM.
In other words,
when non-Black MSM refuse to partner with Black MSM,
this increases the likelihood that Black MSM will partner with each other.
If Black MSM are segregated from other MSM when it comes to finding sex
partners,
a high prevalence of HIV can become self-perpetuating.
Such segregation can make risky behavior more risky than it would be
in an environment with lower prevalence.
%This argument is supported by Das et al. \cite{das2010},
%who found that reductions in community viral load
%were associated with reductions in HIV infections.

Fortunately, many researchers are examining
where MSM meet their sexual partners.
Perhaps due to stigma or the difficulty of finding other MSM,
MSM are more likely than heterosexuals to meet sex partners in designated
meeting places like bars, websites, or apps
than in public spaces, like at work or in
school \cite{glick2012,jennings2015}.
Recently, researchers in
Mississippi \cite{oster2013},
Houston \cite{fujimoto2013},
Los Angeles \cite{holloway2014},
Hong Kong \cite{leung2015},
Baltimore \cite{brantley2017},
Chicago \cite{young2017}, and
Rhode Island \cite{chan2018}
have constructed affiliation networks in an effort to better understand
the transmission of infectious diseases.
Affiliation networks are bipartite graphs with one type of node representing
people and the other type representing the venues or organizations
those people affiliate with.
By definition, each edge in a bipartite graph connects
two nodes of a different type.  
The affiliation networks in these papers have MSM as one type of node
and venues (either brick-and-mortar or online) as the other;
an individual man is connected to a venue if he socializes there or has met
sexual partners there.

Although affiliation networks are used commonly in network science,
there is as of yet no model for how HIV might spread on an
affiliation network or how the network might evolve in time.
In addition,
affiliation networks have not yet been leveraged to examine
HIV risk disparities between Black and non-Black MSM.
In this paper, we propose a model for the evolution of an affiliation network
over time and for the spread of HIV over that network.
In addition, we introduce an estimator of risk and test
its performance using both simulated and empirical data. 
We also explore the potential utility of affiliation
networks to explain risk disparities.
%Section \ref{sec:model} describes the proposed model and estimator,
%Section \ref{sec:data} describes the data, and we present our conclusions and suggestions for future work in \ref{sec:dis}.

\section{Model}
\label{sec:model}

\subsection{Network Generation}

We start by specifying the model for network generation. Let $N$ be the number of men in the population, and let $M$ be the number of
venues.
For each $i\in\{1,\dotsc,N\}$, person $i$ has parameter vector
$\left(\lambda_i,p_{i1},\dotsc,p_{iM}\right)^\top$. 
The number of his sexual encounters across all venues
follows a Poisson process with rate $\lambda_i$.
For each $j\in\{1,\dotsc,M\}$, any given meeting occurs at venue $j$
with probability $p_{ij}$,
independent of all other sexual encounters,
so that person $i$'s encounters
at venue $j$ follow a Poisson
process with rate $\lambda_i p_{ij}$.

Consider the time interval $(0,t)$.
Let $X_{ij}$ be the number of sexual encounters person $i$ has at venue $j$
in this interval;
then $X_{ij} \sim \text{Poisson}\left(\lambda_i p_{ij} t\right)$,
independent of all $X_{kl}$ where either $i \ne k$ or $j \ne l$.
Further, conditional on $X_{ij}$,
the times of person $i$'s encounters at venue $j$
%(of which there are $X_{ij}$)
are distributed independently and uniformly on $(0,t)$.
%That is,
%$T_{ij1},\dotsc,T_{ijX_{ij}} | X_{ij} \overset{\text{iid}}{\sim}\text{uniform}(0,t)$.
Because the individual Poisson processes are independent of each other,
\[
\sum_{i=1}^N X_{ij} \sim \text{Poisson}\left(t\sum_{i=1}^N \lambda_i p_{ij}\right) \text{,}
\]
and conditional on $\sum_{i=1}^N X_{ij}$,
the times of these encounters are distributed independently and uniformly
on $(0,t)$.

Of course, given that we are interested in viral transmission between people, these encounters need to be linked in some way.
We establish this linkage based on the timing of encounters and assume that the first encounter at venue $j$
is linked to the second encounter at that same venue,
the third is linked to the fourth, and so on.
Three potential problems are immediately apparent with this assumption.
First, a person could be linked to himself if two consecutive meetings
at a single venue belong to the same person.
However, if person $i$ has an encounter at time $t_i$,
the probability that he has another encounter at the same venue
before another person is
\[
\frac{\lambda_i p_{ij}}{\sum_k \lambda_k p_{kj}} \to 0
\]
if $\sum_{k} \lambda_k p_{kj}\to\infty$
as $N\to\infty$.
It is reasonable to assume that $\sum_{k} \lambda_k p_{kj}\to\infty$
as $N\to\infty$ because the parameter vectors
$\left(\lambda_k,p_{k1},\dotsc,p_{kM}\right)^\top$
are assumed i.i.d.
%However, if $(a,b)\subset(0,t)$,
%then the probability of at least two encounters from person $i$
%and zero encounters from anyone else occurring in this interval is
%\[
%\left(1 - e^{-\lambda_i p_{ij} (b-a)} - e^{-\lambda_i p_{ij} (b-a)}\lambda_i p_{ij} (b-a)\right) e^{-(b-a)\sum_{k\ne i} \lambda_k p_{kj}} \to 0
%\]
%if $\sum_{k\ne i} \lambda_k p_{kj}\to\infty$
%as $N\to\infty$.
%It is reasonable to assume that $\sum_{k\ne i} \lambda_k p_{kj}\to\infty$
%as $N\to\infty$ because the parameter vectors
%$\left(\lambda_i,p_{i1},\dotsc,p_{iM}\right)^\top$
%are assumed i.i.d.
Thus, with large enough $N$ such events will occur with vanishing
frequency.
Second, two linked meetings could be consecutive but still so far apart
in time that it is unlikely that they correspond to the same event, i.e., an encounter between two individuals. 
Again, if $\sum_{k} \lambda_k p_{kj}\to\infty$
as $N\to\infty$ then the probability of zero events in any interval
approaches zero.
So, with large enough $N$, the meetings will be dense
enough in the interval that consecutive meetings will be reasonably
close in time.
Finally, the last meeting may be ``orphaned'' if the total number of
meetings in the interval is odd.
For parsimony, if this occurs we discard the final meeting.

This process generates both a bipartite affiliation network
and a venue-to-venue network.
For the affiliation network, person $i$ is connected to venue $j$
if $X_{ij} \ge 1$, and the weight of this edge is equal to $X_{ij}$.
For the venue-to-venue network,
venue $j$ is connected to venue $j'$
if at least one participant met a sex partner at both venues,
and the weight of this edge is equal to
the number of participants the two venues share.

\subsection{HIV Transmission}

We next specify the model for HIV transmission. For each $i\in\{1,\dotsc,N\}$, let $Y_{i0} = 1$ if person $i$ is HIV-positive
at time $0$ and let $Y_{i0} = 0$ otherwise;
similarly, let $Y_{it} = 1$ if person $i$ is HIV-positive
at time $t$ and let $Y_{it} = 0$ otherwise.
We assume that $Y_{i0}$ is known,
and we would like to predict $Y_{it}$ for $t>0$, i.e., the HIV status of person $i$ at some later time.
Consider the scenario where an individual who is HIV-negative at time $0$
contracts HIV
independently and with probability $\pi$ for each encounter in $(0,t)$
with a sex partner who is HIV-positive at time $0$.
Clearly this is a simplification.
First, the probability of transmission of HIV is not the same
for each encounter and depends on such things as the sex act
and type of protection used (e.g., condoms, pre-exposure prophylaxis).
Second, if person $i$ is HIV-negative at time $0$ but
becomes infected at time $t' \in (0,t)$,
it would be reasonable to allow person $i$ to transmit the virus to others
in the interval $(t',t)$.
For parsimony, however, we consider the simpler version.

Assume $X_{1j},\dotsc,X_{Nj},Y_{10},\dotsc,Y_{N0}$ are known.
Of the $\sum_{i=1}^N X_{ij}$ encounters at venue $j$ in $(0,t)$,
$\sum_{i=1}^N X_{ij}Y_{i0}$ belong to HIV-positive individuals.
Because these encounters are independently and uniformly distributed
throughout the time interval, each encounter has the same probability of belonging
to an HIV-positive individual.
Denote this probability
\[
Q_j = \frac{\sum_{i=1}^N X_{ij}Y_{i0}}{\sum_{i=1}^N X_{ij}} \text{.}
\]
Thus, if we consider person $i\in\{1,\dotsc,N\}$,
the number of HIV-positive partners he has at venue $j$ in the interval is approximately
Binomial$\left(X_{ij},Q_j\right)$,
and his risk (probability) of contracting HIV is
\[
R_i = 1 - \prod_{j=1}^{M} \left[(1-\pi)Q_j + 1 - Q_j\right]^{X_{ij}} \text{.} 
\]

Note that $\pi$, the probability of per-encounter transmission, is not specific to any individual but instead is shared by all members of the population.

\subsection{Risk Estimator}

In practice, it is unrealistic to assume knowledge of the entire population
of $N$ individuals and their person-site encounters $X_{ij}$,
and instead any estimator of risk needs to be based on a sample.
Let $n$ be the number of individuals in the sample,
and let $m$ be the number of venues reported by participants in the sample.
Assume each individual $i\in\{1,\dotsc,n\}$ was followed in the interval $(-t,0)$.
%(Note that this is an abuse of notation.
%Person $i$ in the population may not correspond to person $i$ in the sample.)
For each venue $j\in\{1,\dotsc,m\}$,
%(the same abuse of notation),
let $Z_{ij}$ be the number of sexual encounters person $i$ had at venue $j$
in the interval $(-t,0)$. We assume that these numbers are known without error. 
Then $\forall j\in\{1,\dotsc,m\}$,
the probability of having an encounter with an HIV-positive individual
at site $j$ can be estimated as 
\[
\hat{Q}_j = \frac{\sum_{i=1}^n Z_{ij}Y_{i0}}{\sum_{i=1}^n Z_{ij}} \text{,}
\]
and $\forall i\in\{1,\dotsc,n\}$, and the risk (probability) of contracting HIV can be estimated as 
\[
\hat{R}_i = 1 - \prod_{j=1}^{m} \left[(1 - \pi) \hat{Q}_j + 1 - \hat{Q}_j\right]^{Z_{ij}} \text{.}
\]
Note that we assume $\pi$ to be known.

\section{Empirical Study}

Among the affiliation network studies listed in the introduction,
only one, the uConnect Study in Chicago
\cite{young2017,lancki2018},
was longitudinal with a mix of HIV-positive
and HIV-negative participants at baseline.
It recruited Black MSM and
collected information on categories of sex partner meeting places
instead of specific, identifiable venues.
As such, the data are not a perfect fit to the model,
and we would not expect the new estimator to outperform
other predictors of risk in this setting.
Still,
this data set contains vital information on the distribution of
total number of sex partners per person
and the distribution of sex partner meeting places per person.
It also serves as a contrast to later simulations
that assume data collection proceeded as dictated by the model.

\subsection{Data}
\label{sec:data}

For Wave 1 of the study,
which occurred between June 2013 and July 2014,
618 African-American MSM between the ages of 16 and 29
were recruited in Chicago
via respondent-driven sampling.
Sixty-five of the participants were recruited independently,
and any participant could recruit up to six other eligible MSM
to participate in the study.
For Wave 2, which occurred between April 2014 and May 2015,
524 of these men were interviewed again.

In the Wave 1 interview,
participants were asked how many people they had had sex with
in the last six months.
For up to six of these sex partners,
starting with the most recent and working backward,
participants were asked 
how they met and how many times they had sex.
Specifically,
the wording of the meeting question was,
``How did you meet [NAME] leading up to the first time you had sex?
Was that through somebody else you both knew,
through a phone or internet program or site, or some other way?''
If the respondent said phone or internet,
the interviewer followed-up with,
``Was that a mobile app, something on the internet or a phone service?''
If the respondent said, ``Some other way,''
the interviewer followed-up with,
``Where did you meet [NAME] for the first time?
Was that at a . . . bar/night club/dance club;
social service or volunteer event;
health club or gym;
private (house) party;
outdoors/cruising/parks/public/bathrooms;
work;
school;
church or house of worship/church or religious activity;
jail or prison;
AA or NA;
other (SPECIFY)''.
The men were not asked about specific locations where they met their
sex partners, just categories.
Section \ref{sec:supplement} contains an alternative analysis
based on where the participants met or socialized with other men,
and includes geographic information.

At Wave 1, there were 1,593 sex partners in the dataset.
At each wave,
participants received HIV tests and were asked about their HIV status.
For the present analysis, HIV status was determined by lab results,
unless those results were missing, in which case HIV status was determined
by self-report.

In the Wave 1 data, 110 participants (18\%) were missing lab HIV  results
and 43 (7\%) were missing self-reported HIV status.
Eight (1\%) were missing both, and they were removed from the data set;
this corresponded to a removal of 14 sex partners (1\%).
Eighteen participants (3\%) were removed because they
did not have any information about their sex partners.
Nine participants (1\%; 93 sex partners, 6\%) were removed because they only
had information about sex partners from more than six months prior
to Wave 1;
three participants (0.5\%; 54 sex partners, 3\%)
were removed because they were missing data
on how they met their sex partners;
109 participants (18\%; 468 sex partners, 29\%)
were removed because they met all their sex partners
``through somebody else'' or ``knew each other previously'';
three participants (0.5\%; eight sex partners, 0.5\%)
were removed because they met their partners through
``phone or internet'' but did not specify whether that was through
a phone service, website, or mobile app;
and two participants (0.3\%; six sex partners, 0.4\%)
were removed because they did not specify
how many times they had had sex with their partners.
This left 466 participants (75\% of the total)
with information on 950 sex partners (60\% of the total)
met at 15 ``venues''.
A list of venues and the number of sex partners met at each is
in Table \ref{table:venues}.

\begin{table}[h!]
\begin{tabular}{l r r}
\toprule
\textbf{Venue} & \textbf{Number of Partners Met There} & \textbf{Cumulative \%} \\
\midrule
Internet site & 265 & 27.9 \\
Mobile app & 190 & 47.9 \\
Outdoors/cruising/parks/public/bathrooms & 163 & 65.1 \\
School & 72 & 72.6 \\
Phone service & 56 & 78.5 \\
Bar/night club/dance club & 41 & 82.8 \\
Ball/dance group/social event & 41 & 87.2 \\
Private (house) party & 38 & 91.2 \\
Work & 30 & 94.3 \\
Boystown & 13 & 95.7 \\
Program/support group & 10 & 96.7 \\
Sex party & 10 & 97.8 \\
Some other way & 9 & 98.7 \\
Church/house of worship/religious activity & 7 & 99.5 \\
Institution & 5 & 100.0 \\
\bottomrule
\end{tabular}
\caption{List of the 15 venues in the final data set
and the number of partners met at each venue.}
\label{table:venues}
\end{table}

One participant had the same date for his Wave 1 and Wave 2 interviews,
and this date was neither the latest date for Wave 1 nor
the earliest date for Wave 2.
His Wave 2 data were deleted.
This left 395 participants with HIV status at Wave 2.
The median elapsed time between Wave 1 and Wave 2 was 266 days.
Of the 288 participants who were HIV-negative at Wave 1,
234 had HIV status data at Wave 2.

Many of the participants reported that they had had sex with
more people over the last six months than they were asked about in detail.
In order to estimate the number $Z_{ij}$ in our risk calculation, 
the number of times person $i$ reported having had
sex with someone he met at site $j$ 
%(across all the people he was asked about in the study)
was multiplied by the number of people he reported having had sex with
over the last six months and divided by the number of sex partners
he had in the data set.
This assumes that the partners a participant was asked about are
exactly representative of his partners over the past six months.

\subsection{Methods}

\begin{figure}
\centering
\includegraphics[width = 0.8\linewidth]{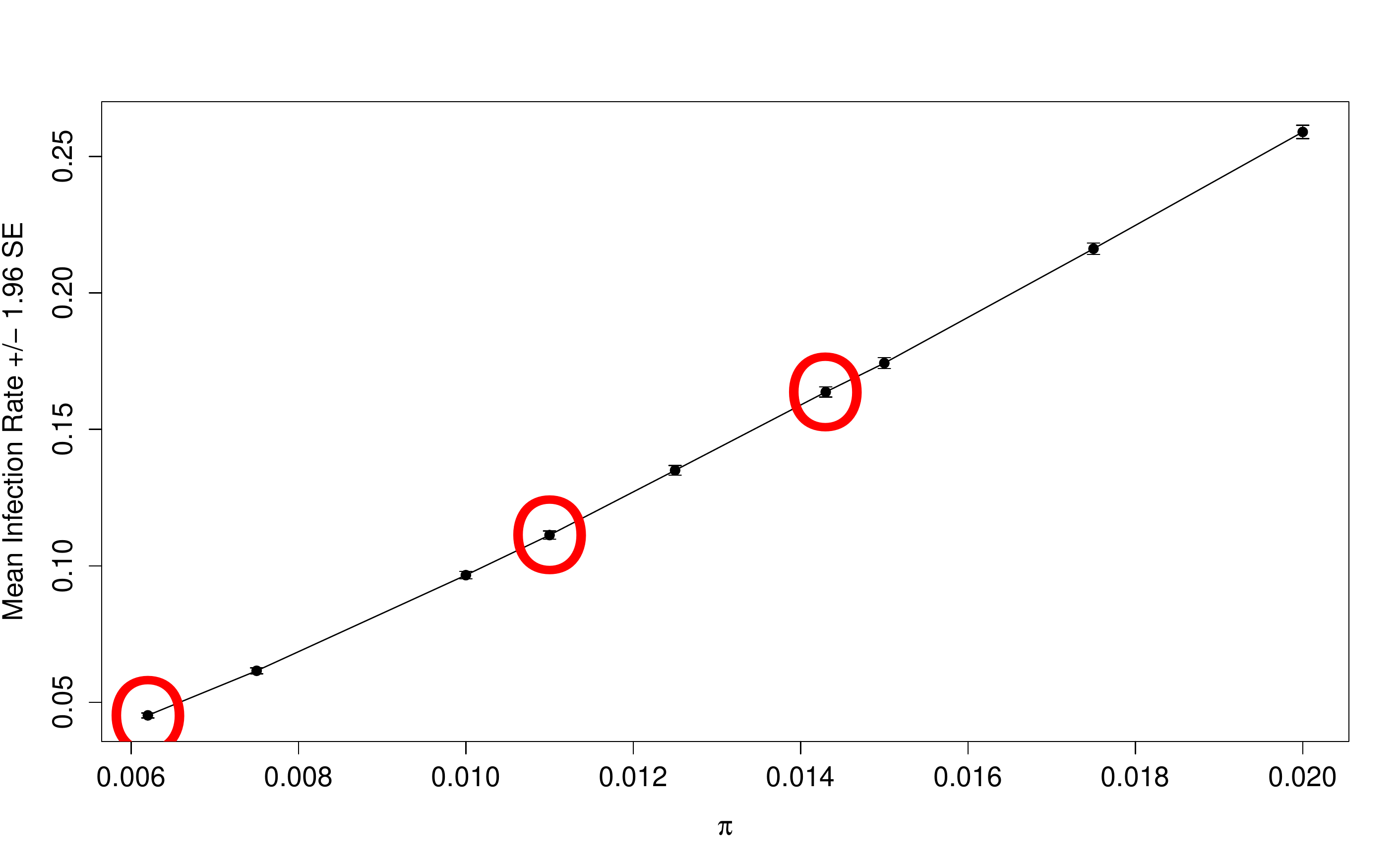}
\caption{\textbf{Mean infection rate as a function of $\pi$.}
These values were obtained by simulating the model described
in Section \ref{sec:model} and varying the
per-contact probability of HIV transmission.
The underlying data are described in Section \ref{sec:data}.
The circles correspond to the values of $\pi$ chosen
for the simulations and to calculate the new predictor of risk.}
\label{figure:infectionrate}
\end{figure}

We calculated five predictors of risk using data for the 15 venues
included in the final data:
\begin{enumerate}
\item Multiple logistic regression based on Wave 1 HIV status:
\[
\logit E\left(Y_{i0}\right)   = \beta_0 + \beta_1 Z_{i1} + \cdots + \beta_{15} Z_{i15} \text{.}
\]
\item Multiple logistic regression based on Wave 2 HIV status:
\[
\logit E\left(Y_{it}\right) = \beta_0 + \beta_1 Z_{i1} + \cdots + \beta_{15} Z_{i15} \text{,}
\]
where $t$ indicates Wave 2.
\item Simple logistic regression based on Wave 1 HIV status:
\[
\logit E\left(Y_{i0}\right) = \beta_0 + \beta_1 \left(Z_{i1} + \cdots +  Z_{i15}\right) \text{.}
\]
\item Simple logistic regression based on Wave 2 HIV status:
\[
\logit E\left(Y_{it}\right) = \beta_0 + \beta_1 \left(Z_{i1} + \cdots + Z_{i15} \right) \text{,}
\]
where $t$ indicates Wave 2.
\item The new method.
$\hat{Q}_j$, $1 \le j \le 15$, was calculated as described above,
and for each of the 234 participants who were HIV-negative at Wave 1
and had HIV status data at Wave 2,
$\hat{R}_i$ was calculated.
Three values of $\pi$ were tested:
$0.62\%$, $1.1\%$, and $1.43\%$.
Two of the values chosen for $\pi$ ($0.62\%$ and $1.43\%$)   
were selected from \cite{jin2010}.
The lower, $0.62\%$, corresponds to the probability of transmission
for insertive unprotected anal intercourse (UAI)
in uncircumcised men.
It was the second-lowest transmission rate reported in \cite{jin2010};
the lowest, $0.11\%$, led to replications with
no new infections in the simulations described below.
The upper value, $1.43\%$,
corresponds to the probability of transmission
for receptive UAI if ejaculation occurred inside the rectum.
It was the highest transmission rate reported in \cite{jin2010}.
The third value chosen for $\pi$ ($1.1\%$)
was chosen because it led to an average of $0.1113$
new infections per 100 person-years at-risk,
close to the value of $0.0974$
that was observed in the data.
See Figure \ref{figure:infectionrate}.
\end{enumerate}
Estimators 1, 3, and 5,
which trained on Wave 1 HIV status,
used data from all 466 participants;
estimators 2 and 4,
which trained on Wave 2 HIV status,
used data from the 395 participants who had Wave 2 HIV status.
The estimators trained on Wave 2 HIV status are intended to
give an upper bound to performance.
Given that they are based on future knowledge,
which would never be available to an investigator or clinician
trying to estimate risk for a patient,
they are not truly fair comparators.
For each estimator,
the AUC was calculated using the 234 participants who were
HIV-negative at Wave 1 and had HIV status at Wave 2.

Summary statistics were calculated for the number of partners
reported by the 466 participants.
%Note that participants were asked about all sex partners,
%including partners who were not cisgender men.
Summary statistics were also calculated for the estimated number of times
participants reported that they had had sex over the previous six months.
Using the notation of the previous section, for participant $i$,
this value would be $\sum_{j=1}^{15} Z_{ij}$.

HIV incidence was estimated using the method described in \cite{neaigus2012}.
The number of new infections between Wave 1 and Wave 2 was the numerator
and the total number of days at risk was the denominator.
This was converted into number of new infections per 100 person-years at-risk.
For those testing negative at both Wave 1 and Wave 2,
the number of days at-risk was the number of days between their
Wave 1 and Wave 2 interviews.
For those testing negative at Wave 1 but positive at Wave 2,
the number of days at-risk was half the number of days between their
Wave 1 and Wave 2 interviews.

\subsection{Results}

Participants reported having a median of 3 sex partners over the past
six months (first quartile: 2; third quartile: 5).
The median number of times they were estimated to have had sex over
the past six months was 18 (first quartile: 8; third quartile: 30).

The AUC for model  1
(multiple logistic regression, Wave 1 outcome) was 0.5695;
the AUC for model 2
(multiple logistic regression, Wave 2 outcome) was 0.5825;
the AUC for model 3
(simple logistic regression, Wave 1 outcome) was 0.3978;
and the AUC for model 4
(simple logistic regression, Wave 2 outcome) was 0.3978.
Whether $\pi$ was set to $0.0062$, $0.0110$, or $0.0143$,
predictor 5 (the new method) yielded an AUC of 0.4256.
Among the 234 participants who were HIV-negative at Wave 1
and had HIV status data at Wave 2,
seventeen tested positive at Wave 2.
There was a total of 63,776.5 person-days at risk,
yielding 0.0974 new infections per 100 person-years at-risk.

\subsection{Discussion}

The new risk estimator performs better than simple logistic regression
using estimated number of sexual encounters as the single
independent variable.
Unfortunately, the new risk estimator performs worse than both chance
and multiple logistic regression.

There are a number of potential reasons that this is the case.
First, the venues listed are not actual venues but categories.
%The most frequently cited venues (internet site,
%mobile app, and outdoors/cruising/parks/public/bathrooms)
They comprise many possible places where the men in the study could meet
their sex partners.
Two men who meet their sex partners exclusively on the internet
could be doing so through two completely different websites.
So, the data do not exactly map onto the proposed model.
Second,
men were asked about their sexual activity over the past six months.
This is a long window of time,
and even people with good memories are bound to make errors
in estimating the number of times they had sex with a given person
during that time period,
or where they met.
In other words, there is some measurement error.
Third,
a large number of participants met their partners through friends
or already knew them,
and these partners were not considered as potential sources of transmission.
Fourth,
the model assumes a constant probability of transmission
for all serodiscordant couplings.
In reality, some couplings will involve condoms, some will involve
pre-exposure prophylaxis (PrEP),
and some will be unprotected.
Further mismatches between the model and the data include
the sampling mechanism (the participants in the dataset were recruited
through a version of respondent-driven sampling
and were not selected uniformly at random from the population of
Black MSM in Chicago);
variable follow-up time across participants;
and follow-up time (approximately nine months) not equaling
the amount of time before Wave 1 for which participants were asked
about their sex partners (six months).

\begin{figure}
\centering
\includegraphics[width = 0.8\linewidth]{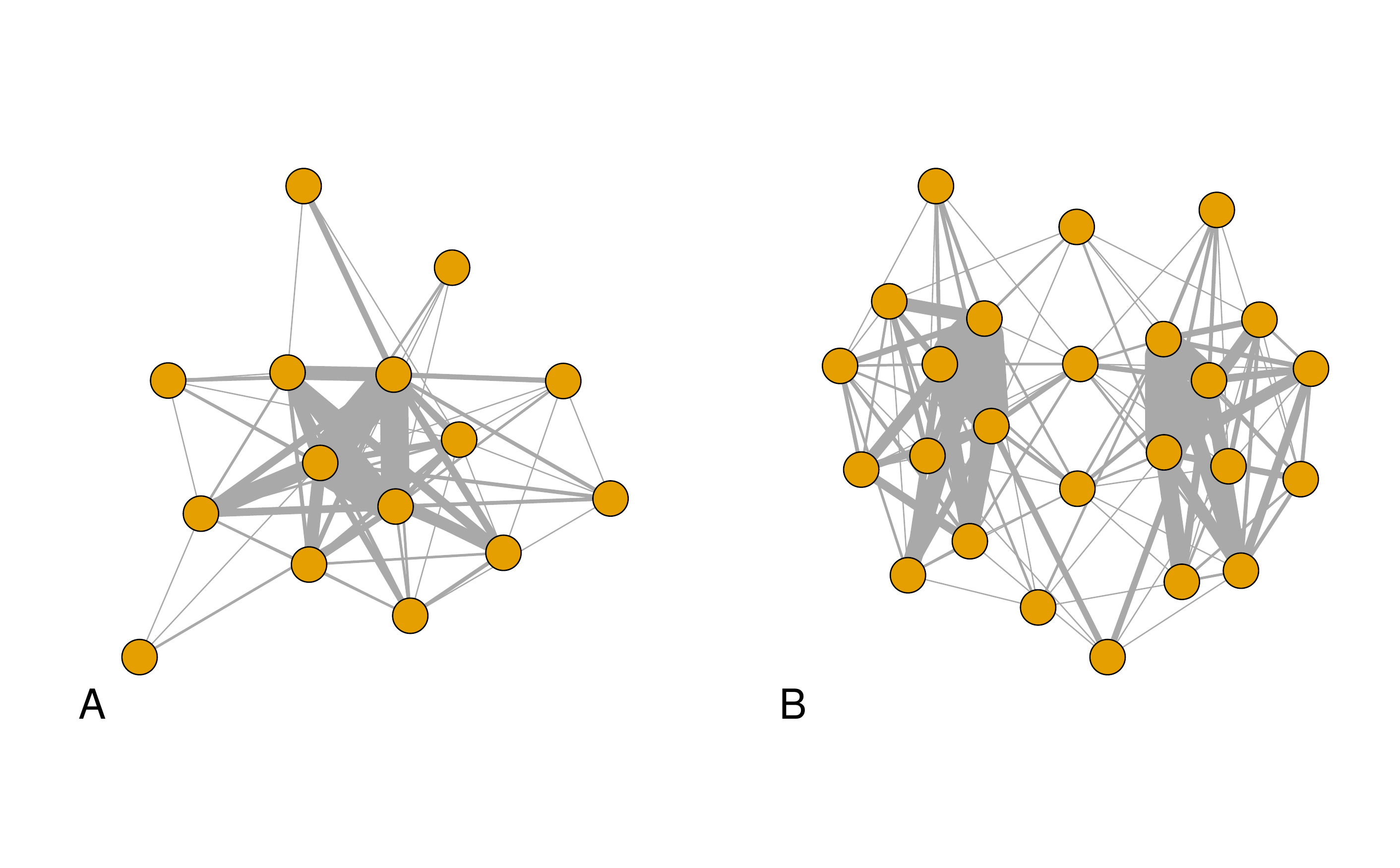}
\caption{\textbf{Venue-to-venue network.}
Each node is a venue,
and two venues are connected by an edge if at least one participant
met a sex partner at both venues.
The width of the edge between two venues corresponds to the number
of participants they share.
(A) The original data set.
(B) The modified data set used in Simulation Study 2.}
\label{figure:venue}
\end{figure}

Figure \ref{figure:venue}A demonstrates another explanation
for the poor performance of the new risk estimator.
It depicts the venue-to-venue graph,
in which each node is a venue
and two venues are connected by an edge if at least one participant
met a sex partner at both venues.
The width of the edge between two venues corresponds to the number
of participants they share.
It's apparent from the figure that the venues are connected in one
overarching cluster.
The participants in the study have a lot of overlap in where they meet
their partners,
so they all have similar risks of contracting HIV.

Although the results are somewhat disappointing,
our simulation approach is able to identify potential reasons for that,
and there are actually good reasons to believe that the method
might perform well in practice if the relevant data were available.

\section{Simulation Study 1}
\label{sec:sim1}

We used simulation to evaluate the performance of the new method
when the model is correct.
As explained below,
we tested the risk estimator when data collection is perfect;
when venues are reported as grouped,
even though they are distinct for the purpose of data generation;
when some venues are not reported;
and when venues are reported incorrectly.

\subsection{Methods}

Sample size $n$ was set to 466;
population size $N$ was set to $n\times 5 =$ 2,330;
the population number of venues $M$ was set to 15;
and $\pi$ was set first to 0.0062,
then to 0.0110,
and then to 0.0143.
There were 1,000 replications for each value of $\pi$.
If we consider each parameter vector to be
$\left(\lambda_i p_{i1}t,\dotsc,\lambda_i p_{iM}t,Y_{i0}\right)^\top$,
then for each replication, the $N$ parameter vectors
were sampled uniformly at random with replacement from the 466
participants in the uConnect study.
That is, each vector
$\left(Z_{i1},\dotsc,Z_{i15},Y_{i0}\right)^\top$
from the previous section was considered to be a potential parameter vector
$\left(\lambda_i p_{i1}t,\dotsc,\lambda_i p_{i15}t,Y_{i0}\right)^\top$
for the simulation study.
Since the participants in the original study were asked about
their sexual activity over the previous six months,
$t$ was set to six months.

Each replication consisted of the following steps:
\begin{enumerate}
\item Draw the $N$ parameter vectors.
\item Simulate the model for six months.
\item Record $Z_{ij}$ and $Y_{i0}$.
\item Draw a sample of size $n$.
\item Calculate the $\hat{Q}_j$ and $\hat{R}_i$ based on the sample.
\item Simulate the model for six more months.
\item Test $\hat{R}_i$ and the four logistic regressions from the previous
section as predictors of $Y_{it}$ for the participants in the sample.
Also,
measure the number of new infections per 100 person-years at-risk.
\end{enumerate}

In addition to varying $\pi$,
we varied the missingness with regards to the venues.
That is, we tested the following scenarios:
\begin{enumerate}
\item Perfect sampling.
In this scenario,
no venues were intentionally excluded
(although a venue could have been excluded from the sample
if none of its patrons were sampled).
\item Coarse sampling.
This scenario was intended to represent participants grouping
different venues into categories instead of reporting them as separate.
The venues were first ordered from most to least patronized
(by $\sum_{i=1}^N Z_{ij}$);
then, the first through third were considered one venue
for the purpose of calculating $\hat{Q}$;
the fourth through sixth were considered one venue;
etc.
In other words, we used
\begin{align*}
\hat{Q}_1 &= \frac{\sum_{j=1}^{3} \sum_{i=1}^n Z_{ij} Y_{i0}}{\sum_{j=1}^{3} \sum_{i=1}^n Z_{ij}} \text{,} \\
\hat{Q}_2 &= \frac{\sum_{j=4}^{6} \sum_{i=1}^n Z_{ij} Y_{i0}}{\sum_{j=4}^{6} \sum_{i=1}^n Z_{ij}} \text{,} \\
\end{align*}
etc.
\item Smallest venues missing.
This scenario was intended to represent participants not reporting
the smallest venues.
For all five risk prediction methods,
the three least-patronized venues were ignored.
\item Largest venues missing.
This scenario was intended to represent participants not reporting
the venues that led to the highest numbers of sexual contacts.
For all five risk prediction methods,
the three venues with the highest values of $\sum_{i=1}^n Z_{ij}$
were ignored.
\item Contaminated reporting.
This scenario was intended to represent participants reporting
the wrong venues.
For the purpose of calculating all five risk estimators,
50\% of person $i$'s encounters at each venue were redistributed
uniformly at random across all venues.
\end{enumerate}

\subsection{Results}

Across 1,000 replications,
the mean first, second, and third quartiles
of the number of encounters per person
were $7$, $15$, and $27$, respectively.
These values are taken only from the simulation with $\pi = 0.0062$,
but note that the value of $\pi$ does not affect the number of encounters
per person.

\begin{figure}[h!]
\centering
\includegraphics[width = 0.8\linewidth]{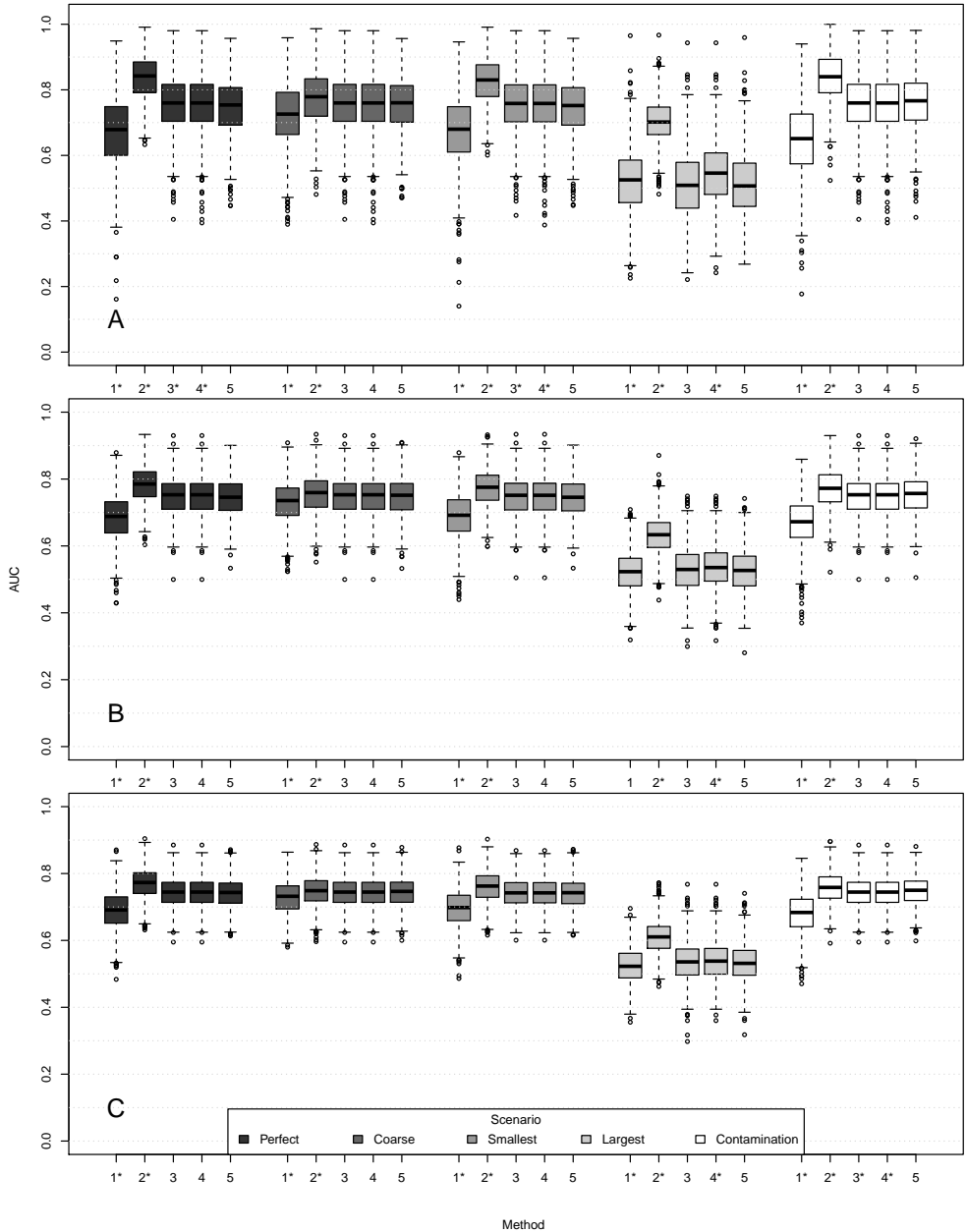}
\caption{\textbf{AUCs from 1,000 replications, single cluster.}
1 = Multiple logistic regression trained on Wave 1 outcome;
2 = Multiple logistic regression trained on Wave 2 outcome;
3 = Simple logistic regression trained on Wave 1 outcome;
4 = Simple logistic regression trained on Wave 2 outcome;
5 = New method.
An asterisk indicates that that method was significantly different
from the new method at the 0.05 level.
In panel (A) $\pi = 0.0062$, 
in (B) $\pi = 0.0110$, and 
in (C) $\pi = 0.0143$.}
\label{figure:boxplota}
\end{figure}

AUCs are presented in Figure \ref{figure:boxplota}.
Figure \ref{figure:boxplota}A displays the AUCs for $\pi = 0.0062$,
Figure \ref{figure:boxplota}B displays the AUCs for $\pi = 0.0110$, and
Figure \ref{figure:boxplota}C displays the AUCs for $\pi = 0.0143$.
In general, as $\pi$ increases, the variability of the performance decreases.

For each value of $\pi$,
and for the Perfect, Coarse, Smallest, and Contamination scenarios,
multiple logistic regression trained on the Wave 1 outcome performs worst;
simple logistic regression performs about the same as
or slightly better than the new method,
regardless of whether it is trained on the Wave 1 or Wave 2 outcome;
and multiple logistic regression trained on the Wave 2 outcome performs best.
For the Largest scenario,
simple logistic regression trained on the Wave 1 outcome performs
about the same as the new method;
multiple logistic regression trained on the Wave 1 outcome
performs about as well as simple logistic regression trained on
the Wave 2 outcome;
and multiple logistic regression trained on the Wave 2 outcome performs
the best.
In this missing data scenario,
the AUCs for all methods are much lower than for other missing data scenarios.

\subsection{Discussion}

As $\pi$ decreases, new infections become more rare,
making the relationship between the venues and the outcome
more dependent on chance.
This explains the increasing variability of the risk estimators
with decreasing values of $\pi$.

The similarity of the performance of the new method
with the simple logistic regressions seems to indicate that
an individual's pattern of venue visitation is not as important
as his total number of sexual encounters.
This is corroborated by the performance of the multiple logistic regressions,
which seem to overfit the outcome to the data.
This overfitting causes the multiple logistic regression trained on the
Wave 1 outcome to display the worst performance
and the multiple logistic regression trained on the Wave 2 outcome
to display the best performance.
Another piece of evidence for the primacy of number of sexual encounters
is the Contamination scenario.
Here, the reported pattern of venue visitation differs greatly
from the true pattern of venue visitation,
but each person's total number of encounters remains the same.
The performance of each method is about the same in this scenario
as in the Perfect scenario,
indicating that the total number of encounters is what matters.
The Largest scenario provides a contrast to the Contamination scenario;
here, the three venues with the most sexual encounters are not reported.
This causes a decrease in the total number of sexual encounters
reported by many participants,
and the performance of all five risk estimators suffers.

The distribution of the number of sexual encounters per person
is approximately the same in the simulation as in the dataset
on which it is based.

\section{Simulation Study 2}
\label{sec:sim2}

The following simulation study was intended to address the
dense clustering of the venues demonstrated in Figure \ref{figure:venue}A.
A new dataset was created, this time with two clusters
and a much lower HIV prevalence in one of the clusters.

\subsection{Methods}

This simulation study was exactly the same as the first,
with three exceptions.
First, the 466 rows of the dataset were duplicated,
generating a new dataset of 932 participants.
For the second group of 466 participants,
the ten largest venues were renamed.
This meant there were 25 total venues,
with the first 466 participants only overlapping with the second 466
participants at the five smallest venues.
The resulting venue-to-venue graph is in Figure \ref{figure:venue}B.
Second, this second group of participants was also modified in that
each participant who was HIV-positive at Wave 1
was changed to be HIV-negative at Wave 1 with probability $0.75$.
As a result,
the second group had an HIV prevalence at Wave 1 of $0.0815$,
whereas the first group had an HIV prevalence at Wave 1 of $0.3820$.
Third, the population size $N$ was set to $932\times 5 = 4,660$.
This new dataset formed the basis for the second simulation study
in that in each replication,
862 participants were drawn uniformly at random from it.

\subsection{Results}

\begin{figure}[h!]
\centering
\includegraphics[width = 0.8\linewidth]{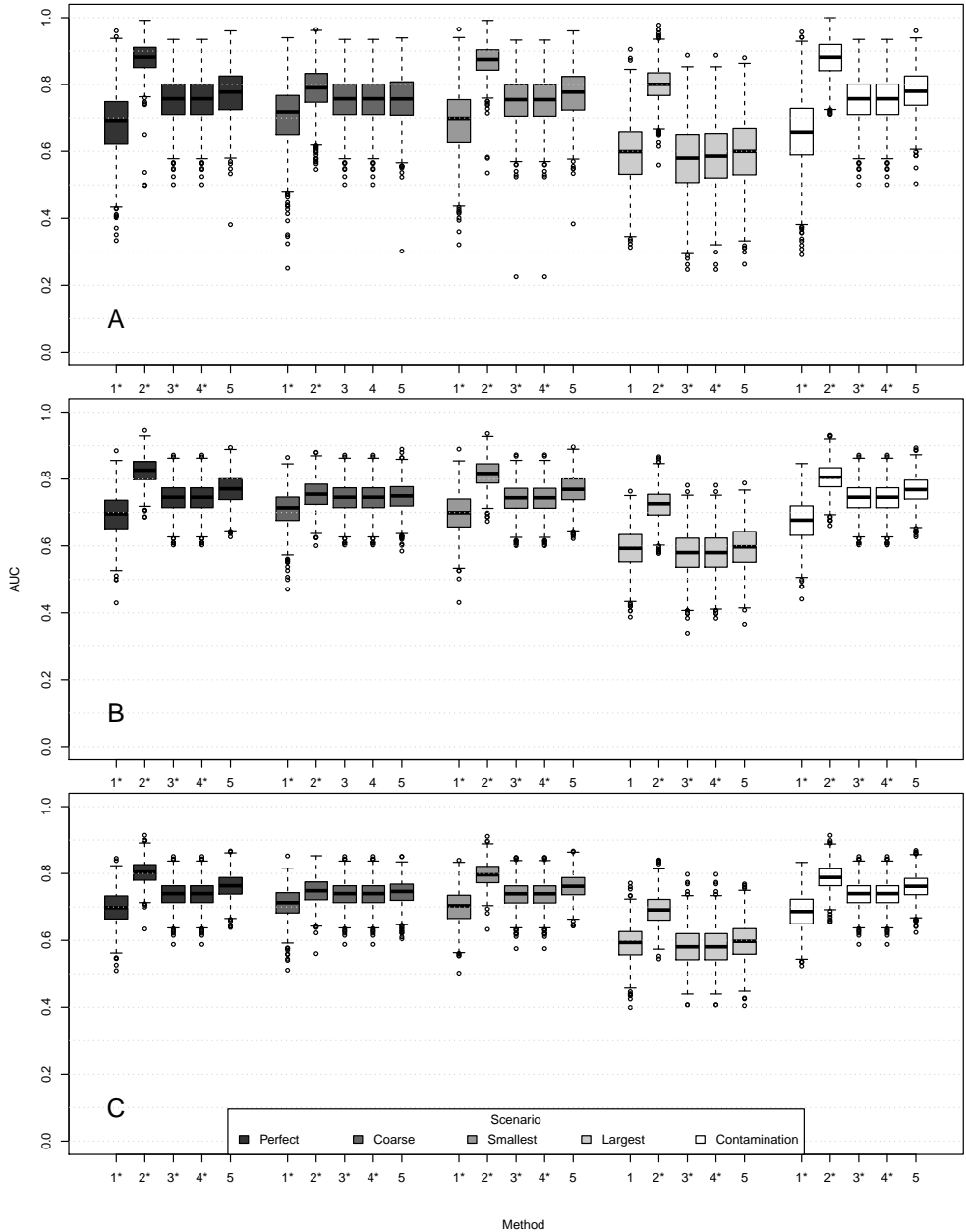}
\caption{\textbf{AUCs from 1,000 replications, two clusters.}
1 = Multiple logistic regression trained on Wave 1 outcome;
2 = Multiple logistic regression trained on Wave 2 outcome;
3 = Simple logistic regression trained on Wave 1 outcome;
4 = Simple logistic regression trained on Wave 2 outcome;
5 = New method. 
An asterisk indicates that that method was significantly different
from the new method at the 0.05 level.
In panel (A) $\pi = 0.0062$, 
in (B) $\pi = 0.0110$, and 
in (C) $\pi = 0.0143$.}
\label{figure:boxplotj}
\end{figure}

AUCs are presented in Figure \ref{figure:boxplotj}.
Figure \ref{figure:boxplotj}A displays the AUCs for $\pi = 0.0062$,
Figure \ref{figure:boxplotj}B displays the AUCs for $\pi = 0.0110$, and
Figure \ref{figure:boxplotj}C displays the AUCs for $\pi = 0.0143$.
In general, as $\pi$ increases, the variability of the performance decreases.

For each value of $\pi$,
and for the Perfect, Coarse, Smallest, and Contamination scenarios,
multiple logistic regression trained on the Wave 1 outcome performs worst;
simple logistic regression performs better;
the new method performs even better;
and multiple logistic regression trained on the Wave 2 outcome performs best.
For the Largest scenario,
the pattern is repeated except that multiple logistic regression
trained on the Wave 1 outcome performs about as well as the new method.
In this missing data scenario,
the AUCs for all methods are much lower than for other missing data scenarios.

\subsection{Discussion}

The pattern of results for the Perfect, Coarse, Smallest,
and Contamination scenarios for this simulation study
differs from that for the previous simulation study
in that the new method performs better than the two
simple logistic regressions.
This seems to indicate that the pattern of venue visitation
is more relevant than the simple total number of sexual encounters per person.
The superiority of the new method over multiple logistic regression
trained on the Wave 1 outcome also indicates that the new method
is not overfitting.

\section{Discussion}
\label{sec:dis}

This paper proposes a new model for HIV transmission based on where
MSM meet their sexual partners,
and tests a measure of risk based on that model in both real-world
and simulated data.
The measure of risk performs about as well as simple logistic regression
when the venues where people meet their sex partners contains one large cluster,
but performs better than both simple and multiple logistic regression
when there are two clusters that differ in their HIV prevalence.

A plurality of the sex partners in the real-world sample
were met through friends or were already known to the participants,
and thus do not fit into the proposed model for HIV transmission.
In order to account for that,
this model could be extended by allowing people to be considered ``venues.''

Admittedly, the assumption that sexual encounters follow a Poisson process
is a strong one.
It means that an individual's rate of sexual encounters does not change
over time,
even if he finds out he has been infected with HIV.
Gorbach, Javanbakht, and Bolan \cite{gorbach2018}
found that the median number of partners among HIV-positive
MSM in Los Angeles did decrease after diagnosis.
That said,
even if the newly-diagnosed men change their behavior,
the model still uses information from men who already knew they were
HIV-positive and from men who remain HIV-negative,
and their behavior would not be expected to change.
Also,
the model is based on sampling from a population of MSM visiting these venues,
so even if the sampled men change their behavior,
the unsampled men may not.
Finally,
Gorbach, Javanbakht, and Bolan \cite{gorbach2018} found that
the rate of condomless anal intercourse (CAI) among HIV-positive
MSM did not decrease after diagnosis.
Further studies may examine how this complex behavior change affects
HIV risk for HIV-negative men in the same sexual network;
the current paper is intended to be a starting point.

There are many ways to enrich this model.
We ignored methods of protection such as condoms and pre-exposure
prophylaxis (PrEP), as well as the fact that an HIV-positive person with
undetectable viral load will not transmit the virus, even during
unprotected sex.
These realities can be incorporated into expansions of the model.
We also ignored temporal patterns like people being more likely to attend
bars and clubs on the weekends.
However, if the interval $(-t,t)$
covers enough weeks,
these temporal patterns may average out.

Although the proposed model was developed for HIV transmission in MSM,
it does not need to be limited to this pathogen or this population.
A natural extension is hepatitis in injection drug users (IDUs).
Any IDU can share a needle with any other IDU without regard to gender,
and IDUs may be restricted to meeting each other through a limited set
of venues or drug dealers.
Instead of sexual encounters, each IDU may have their own rate of injection
drug use.
An affiliation network of IDUs and hotels as sites of injection has already been
created in Winnipeg \cite{wylie2007}.
%One can also consider ties between homeless individuals meeting in shelters
%\cite{desmond2012}.

\section{Declarations}

\subsection{Ethics approval and consent to participate}
As the authors obtained only de-identified data for this study,
it was determined ``not human subjects research''
by the IRB of the Harvard T.H. Chan School of Public Health.

\subsection{Availability of data and materials}
The data that support the findings of this study are available from
the uConnect Study Team \cite{young2017,lancki2018},
but restrictions apply to the availability of these data,
which were used under license for the current study,
and so are not publicly available.

\subsection{Declarations of interest}
None.

\subsection{Funding}
Jonathan Larson is funded by NIH 2T32AI007358-31.
Jukka-Pekka Onnela is funded by NIH R01AI138901.
The uConnect study was funded by NIH R01DA033875.
The funders had no role in study design;
in the collection, analysis and interpretation of data;
in the writing of the report;
or in the decision to submit the article for publication.

\subsection{Authors' contributions}
JL conceptualized and designed the study,
conducted the analysis, and wrote the original draft.
Both authors reviewed, edited, and approved the manuscript.

\subsection{Acknowledgements}
We would like to thank Lindsay Young,
John Schneider, Stuart Michaels, Kayo Fujimoto, Hildie Cohen,
and everyone on the uConnect team for sharing their data.
We are also grateful for suggestions made by
Alessandro Vespignani and Edoardo Airoldi.

\bibliographystyle{vancouver}
\bibliography{bib}

\section{Supplement}
\label{sec:supplement}

\subsection{Empirical Study}

\subsubsection{Data}
\label{sec:data2}

In the Wave 1 interview,
participants were asked how often they had gone to clubs or bars;
gyms; malls, shopping centers, or outdoor or public spaces;
adult book stores or bathhouses; or ball scenes
``to meet or socialize with other men''.
The possible responses were ``Every day'',
``Several times a week'', ``Once a week'',
``Once every two weeks'', ``Once a month'',
``A couple of times a year'',
``Once a year'',
``Less than once a year'', and ``Never''.
For each of these venue types,
participants were asked if they were located in
the South Side, North Side, West Side, East Side,
South Suburbs, or other part of Chicago.
At Waves 1 and 2,
participants received HIV tests and were asked about their HIV status.
For the present analysis, HIV status was determined by lab results,
unless those results were missing, in which case HIV status was determined
by self-report.

In the Wave 1 data, 110 participants (18\%) were missing lab HIV  results
and 43 (7\%) were missing self-reported HIV status.
Eight (1\%) were missing both, and they were removed from the data set.
Four participants (0.6\%) were missing data on where they met or socialized
with other men and were removed from the data set.
One participant had the same date for his Wave 1 and Wave 2 interviews,
and this date was neither the latest date for Wave 1 nor
the earliest date for Wave 2.
His Wave 2 data were deleted.
This left 606 participants at Wave 1 (98.1\% of the total)
and 512 (82.8\%) participants at Wave 2.
Of the participants at Wave 2, 509 had HIV information.

The frequency data were recoded as follows
to reflect number of visits to a venue type (e.g., clubs and bars)
over the course of nine months,
the average elapsed time between Wave 1 and Wave 2.
Thus ``Every day'' became 270;
``Several times a week'' became 116, or three times per week;
``Once a week'' became 39;
``Once every two weeks'' became 19;
``Once a month'' became 9,
``A couple of times a year'' became 4;
``Once a year'' became 1;
and ``Less than once a year'' and ``Never'' became 0.
These values were then divided evenly across the neighborhoods
the participant said he visited.
For example, if a participant said he visited clubs and bars
several times a week,
and these clubs and bars were located on the North and South Sides,
he was coded as visiting clubs and bars on the North Side 58 times
and clubs and bars on the South Side 58 times.
Two participants said they visited clubs and bars but didn't know where,
and one participant said he visited ball scenes but didn't know where.
These participants were not removed from the data set,
but were coded as not visiting clubs and bars or ball scenes,
respectively.
Twenty-seven participants said they visited a venue type but
did not state where.
They were removed from the data set.
This left 579 participants (93.7\% of the total) at Wave 1
and 488 (79.0\%) at Wave 2.

The median elapsed time between Wave 1 and Wave 2 was 267.5 days.
Of the 362 participants who were HIV-negative at Wave 1,
293 had HIV status data at Wave 2.

A list of the venues and the estimated number of visits there
across all participants in a nine-month period
is in Table \ref{table:venuesalt}.

\begin{table}
\centering
\begin{tabular}{l r r}
\toprule
\textbf{Venue} & \textbf{\# Visits} & \textbf{\%} \\
\midrule
Clubs and Bars, North Side & 4,914 & 24.4 \\
Clubs and Bars, East Side & 290 & 1.4 \\
Clubs and Bars, South Side & 1,375 & 6.8 \\
Clubs and Bars, West Side & 237 & 1.2 \\
Clubs and Bars, South Suburbs & 213 & 1.1 \\
Clubs and Bars, Other & 305 & 1.5 \\
Gyms, North Side & 446 & 2.2 \\
Gyms, East Side & 61 & 0.3 \\
Gyms, South Side & 868 & 4.3 \\
Gyms, West Side & 237 & 1.2 \\
Gyms, South Suburbs & 24 & 0.1 \\
Gyms, Other & 98 & 0.5 \\
Public Spaces, North Side & 1,841 & 9.1 \\
Public Spaces, East Side & 407 & 2.0 \\
Public Spaces, South Side & 3,050 & 15.1 \\
Public Spaces, West Side & 1,251 & 6.2 \\
Public Spaces, South Suburbs & 1,087 & 5.4 \\
Public Spaces, Other & 123 & 0.6 \\
Bathhouses and Bookstores, North Side & 542 & 2.7 \\
Bathhouses and Bookstores, East Side & 38 & 0.2 \\
Bathhouses and Bookstores, South Side & 16 & 0.1 \\
Bathhouses and Bookstores, West Side & 291 & 1.4 \\
Bathhouses and Bookstores, South Suburbs & 1 & 0.0 \\
Bathhouses and Bookstores, Other & 21 & 0.1 \\
Balls, North Side & 402 & 2.0 \\
Balls, East Side & 112 & 0.6 \\
Balls, South Side & 1,350 & 6.7 \\
Balls, West Side & 434 & 2.2 \\
Balls, South Suburbs & 14 & 0.1 \\
Balls, Other & 0 & 0.0 \\
\bottomrule
\end{tabular}
\caption{\label{table:venuesalt}
List of the venues in the final data set
and the estimated number of visits to each venue.}
\end{table}

\subsubsection{Methods}

\begin{figure}
\centering
\includegraphics[width = 0.8\linewidth]{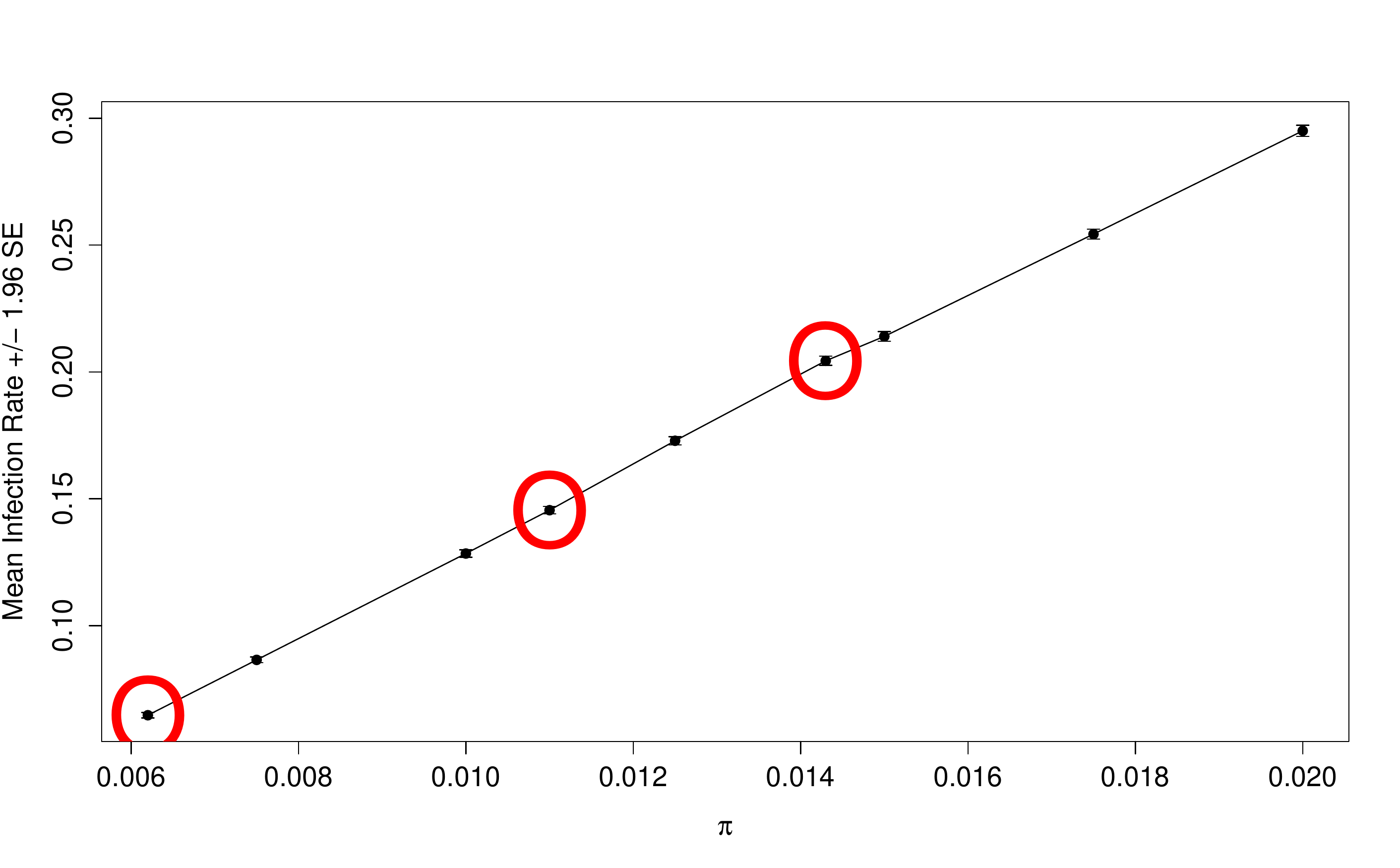}
\caption{\textbf{Mean infection rate as a function of $\pi$.}
These values were obtained by simulating the model described
in Section \ref{sec:model} and varying the
per-contact probability of HIV transmission.
The underlying data are described in Section \ref{sec:data2}.
The circles correspond to the values of $\pi$ chosen
for the simulations and to calculate the new predictor of risk.}
\label{figure:infectionratealt}
\end{figure}

We calculated five predictors of risk using data for the 30 venues
included in the final data:
\begin{enumerate}
\item Multiple logistic regression based on Wave 1 HIV status:
\[
\logit E\left(Y_{i0}\right)   = \beta_0 + \beta_1 Z_{i1} + \cdots + \beta_{30} Z_{i30} \text{.}
\]
\item Multiple logistic regression based on Wave 2 HIV status:
\[
\logit E\left(Y_{it}\right) = \beta_0 + \beta_1 Z_{i1} + \cdots + \beta_{30} Z_{i30} \text{,}
\]
where $t$ indicates Wave 2.
\item Simple logistic regression based on Wave 1 HIV status:
\[
\logit E\left(Y_{i0}\right) = \beta_0 + \beta_1 \left(Z_{i1} + \cdots +  Z_{i30}\right) \text{.}
\]
\item Simple logistic regression based on Wave 2 HIV status:
\[
\logit E\left(Y_{it}\right) = \beta_0 + \beta_1 \left(Z_{i1} + \cdots + Z_{i30} \right) \text{,}
\]
where $t$ indicates Wave 2.
\item The new method.
$\hat{Q}_j$, $1 \le j \le 30$, was calculated as described above,
and for each of the 293 participants who were HIV-negative at Wave 1
and had HIV status data at Wave 2,
$\hat{R}_i$ was calculated.
Three values of $\pi$ were tested:
$0.62\%$, $1.1\%$, and $1.43\%$.
Two of the values chosen for $\pi$ ($0.62\%$ and $1.43\%$)   
were selected from \cite{jin2010}.
The lower, $0.62\%$, corresponds to the probability of transmission
for insertive unprotected anal intercourse (UAI)
in uncircumcised men.
It was the second-lowest transmission rate reported in \cite{jin2010};
the lowest, $0.11\%$, led to replications with
no new infections in the simulations described below.
The upper value, $1.43\%$,
corresponds to the probability of transmission
for receptive UAI if ejaculation occurred inside the rectum.
It was the highest transmission rate reported in \cite{jin2010}.
The third value chosen for $\pi$ ($1.1\%$)
was chosen because it led to an average of $0.1455$
new infections per 100 person-years at-risk,
close to the value of $0.1193$
that was observed in the data.
See Figure \ref{figure:infectionratealt}.
\end{enumerate}
Estimators 1, 3, and 5,
which trained on Wave 1 HIV status,
used data from all 579 participants;
estimators 2 and 4,
which trained on Wave 2 HIV status,
used data from the 485 participants who had Wave 2 HIV status.
The estimators trained on Wave 2 HIV status are intended to
give an upper bound to performance.
Given that they are based on future knowledge,
which would never be available to an investigator or clinician
trying to estimate risk for a patient,
they are not truly fair comparators.
For each estimator,
the AUC was calculated using the 293 participants who were
HIV-negative at Wave 1 and had HIV status at Wave 2.

%Summary statistics were calculated for the number of partners
%reported by the 431 participants.
%Note that participants were asked about all sex partners,
%including partners who were not cisgender men.
Summary statistics were also calculated for the estimated number of times
participants reported that went somewhere to meet or socialize with
other men.
Using the notation of the previous section, for participant $i$,
this value would be $\sum_{j=1}^{15} Z_{ij}$.

HIV incidence was estimated using the method described in \cite{neaigus2012}.
The number of new infections between Wave 1 and Wave 2 was the numerator
and the total number of days at risk was the denominator.
This was converted into number of new infections per 100 person-years at-risk.
For those testing negative at both Wave 1 and Wave 2,
the number of days at-risk was the number of days between their
Wave 1 and Wave 2 interviews.
For those testing negative at Wave 1 but positive at Wave 2,
the number of days at-risk was half the number of days between their
Wave 1 and Wave 2 interviews.

\subsubsection{Results}

Participants reported making a median of 10 visits to one of the 30 venues
in order to meet or socialize with other men
(first quartile: 4; third quartile: 39).

The AUC for model  1
(multiple logistic regression, Wave 1 outcome) was 0.4850;
the AUC for model 2
(multiple logistic regression, Wave 2 outcome) was 0.5990;
the AUC for model 3
(simple logistic regression, Wave 1 outcome) was 0.5889;
and the AUC for model 4
(simple logistic regression, Wave 2 outcome) was 0.5889.
Whether $\pi$ was set to $0.0062$, $0.0110$, or $0.0143$,
predictor 5 (the new method) yielded an AUC of 0.4092.
Among the 293 participants who were HIV-negative at Wave 1
and had HIV status data at Wave 2,
sixteen tested positive at Wave 2.
There was a total of 79,614.5 person-days at risk,
yielding 0.1193 new infections per 100 person-years at-risk.

\subsubsection{Discussion}

The new risk estimator performs worse than both chance
and logistic regression.
There are a number of potential reasons that this is the case.
First, the venues listed are not actual venues but category-neighborhood
combinations.
%The most frequently cited venues (internet site,
%mobile app, and outdoors/cruising/parks/public/bathrooms)
Each comprises multiple possible places where the men in the study could meet
their sex partners.
So, the data do not exactly map onto the proposed model.
Second,
men were asked to estimate how often they visited each category of venue.
This estimate is bound to involve error,
and that error is compounded by the fact that categorical responses
were translated into integer values.
Third,
the participants were asked about where they met or socialized
with other men,
not where they met their sex partners.
Fourth,
the model assumes a constant probability of transmission
for all serodiscordant couplings.
In reality, some couplings will involve condoms, some will involve
pre-exposure prophylaxis (PrEP),
and some will be unprotected.
Further mismatches between the model and the data include
the sampling mechanism (the participants in the dataset were recruited
through a version of respondent-driven sampling
and were not selected uniformly at random from the population of
Black MSM in Chicago);
and variable follow-up time across participants.

\begin{figure}
\centering
\includegraphics[width = 0.8\linewidth]{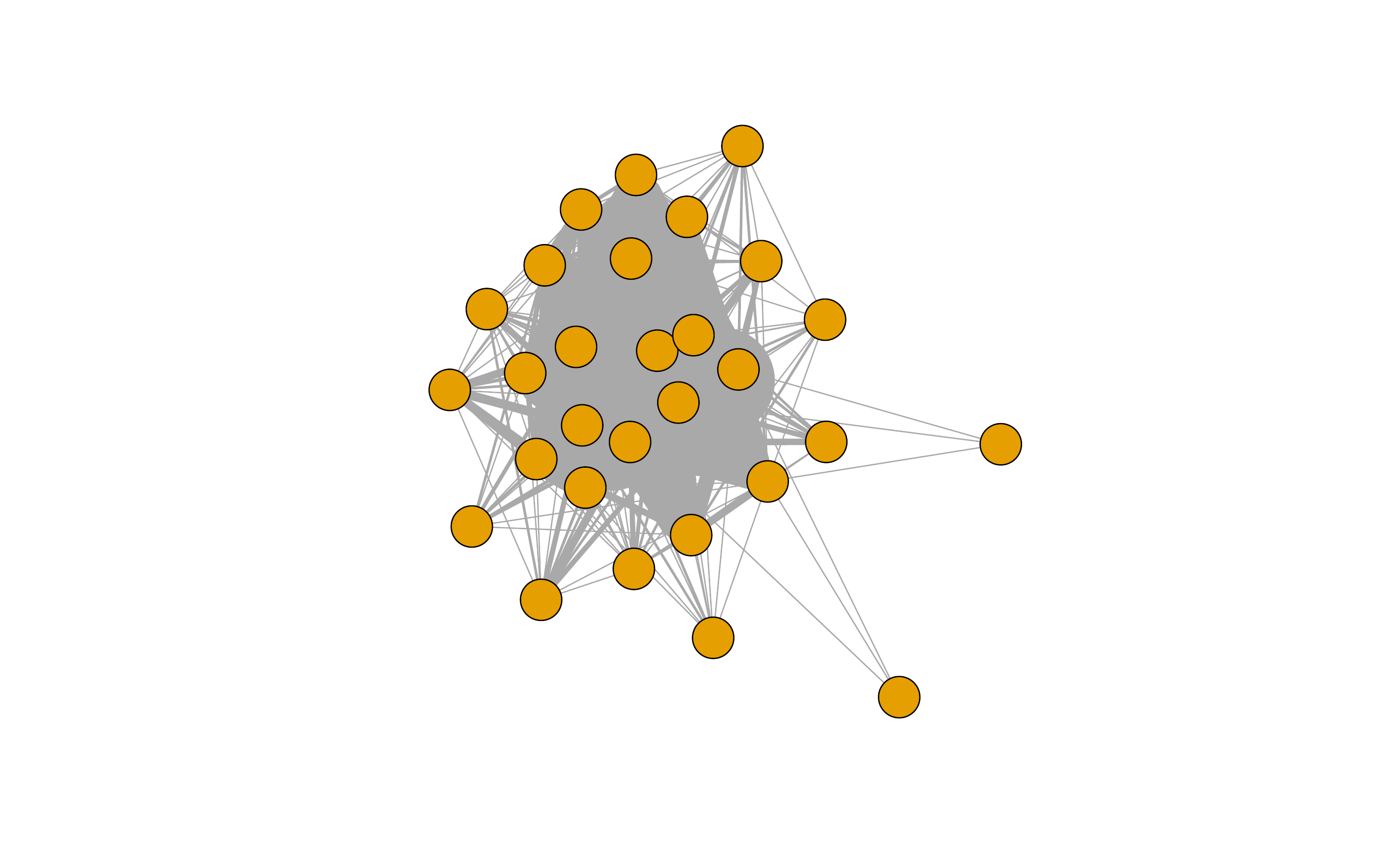}
\caption{\textbf{Venue-to-venue network for the original data.}
Each node is a venue,
and two venues are connected by an edge if at least one participant
visited both.
The width of the edge between two venues corresponds to the number
of participants they share.}
\label{figure:venuealt}
\end{figure}

Figure \ref{figure:venuealt} demonstrates another explanation
for the poor performance of the new risk estimator.
It depicts the venue-to-venue graph,
in which each node is a venue
and two venues are connected by an edge if at least one participant
visited both.
The width of the edge between two venues corresponds to the number
of participants they share.
It's apparent from the figure that the venues are connected in one
overarching cluster.
The participants in the study have a lot of overlap in where they meet
their partners,
so they all have similar risks of contracting HIV.

Although the results are somewhat disappointing,
our simulation approach is able to identify potential reasons for that,
and there are actually good reasons to believe that the method
might perform well in practice if the relevant data were available.

\subsection{Simulation Study 1}

We used simulation to evaluate the performance of the new method
when the model is correct.
What were visits to a venue in the empirical data study
became sexual encounters in this simulation.
As explained below,
we tested the risk estimator when data collection is perfect;
when venues are reported as grouped,
even though they are distinct for the purpose of data generation;
when some venues are not reported;
and when venues are reported incorrectly.

\subsubsection{Methods}

Sample size $n$ was set to 579;
population size $N$ was set to $n\times 5 =$ 2,895;
the population number of venues $M$ was set to 30;
and $\pi$ was set first to 0.0062,
then to 0.0110,
and then to 0.0143.
There were 1,000 replications for each value of $\pi$.
If we consider each parameter vector to be
$\left(\lambda_i p_{i1}t,\dotsc,\lambda_i p_{iM}t,Y_{i0}\right)^\top$,
then for each replication, the $N$ parameter vectors
were sampled uniformly at random with replacement from the 579
participants in the uConnect study.
That is, each vector
$\left(Z_{i1},\dotsc,Z_{i15},Y_{i0}\right)^\top$
from the previous section was considered to be a potential parameter vector
$\left(\lambda_i p_{i1}t,\dotsc,\lambda_i p_{i15}t,Y_{i0}\right)^\top$
for the simulation study.
Since the participants in the original study were asked about
their sexual activity over the previous six months,
$t$ was set to six months.

Each replication consisted of the following steps:
\begin{enumerate}
\item Draw the $N$ parameter vectors.
\item Simulate the model for nine months.
\item Record $Z_{ij}$ and $Y_{i0}$.
\item Draw a sample of size $n$.
\item Calculate the $\hat{Q}_j$ and $\hat{R}_i$ based on the sample.
\item Simulate the model for nine more months.
\item Test $\hat{R}_i$ and the four logistic regressions from the previous
section as predictors of $Y_{it}$ for the participants in the sample.
Also,
measure the number of new infections per 100 person-years at-risk.
\end{enumerate}

In addition to varying $\pi$,
we varied the missingness with regards to the venues.
That is, we tested the following scenarios:
\begin{enumerate}
\item Perfect sampling.
In this scenario,
no venues were intentionally excluded
(although a venue could have been excluded from the sample
if none of its patrons were sampled).
\item Coarse sampling.
This scenario was intended to represent participants grouping
different venues into categories instead of reporting them as separate.
The venues were first ordered from most to least patronized
(by $\sum_{i=1}^N Z_{ij}$);
then, the first through third were considered one venue
for the purpose of calculating $\hat{Q}$;
the fourth through sixth were considered one venue;
etc.
In other words, we used
\begin{align*}
\hat{Q}_1 &= \frac{\sum_{j=1}^{3} \sum_{i=1}^n Z_{ij} Y_{i0}}{\sum_{j=1}^{3} \sum_{i=1}^n Z_{ij}} \text{,} \\
\hat{Q}_2 &= \frac{\sum_{j=4}^{6} \sum_{i=1}^n Z_{ij} Y_{i0}}{\sum_{j=4}^{6} \sum_{i=1}^n Z_{ij}} \text{,} \\
\end{align*}
etc.
\item Smallest venues missing.
This scenario was intended to represent participants not reporting
the smallest venues.
For all five risk prediction methods,
the three least-patronized venues were ignored.
\item Largest venues missing.
This scenario was intended to represent participants not reporting
the venues that led to the highest numbers of sexual contacts.
For all five risk prediction methods,
the three venues with the highest values of $\sum_{i=1}^n Z_{ij}$
were ignored.
\item Contaminated reporting.
This scenario was intended to represent participants reporting
the wrong venues.
For the purpose of calculating all five risk estimators,
50\% of person $i$'s encounters at each venue were redistributed
uniformly at random across all venues.
\end{enumerate}

\subsubsection{Results}

Across 1,000 replications,
the mean first, second, and third quartiles
of the number of visits per person
were $1.9$, $11.2$, and $33.2$, respectively.
These values are taken only from the simulation with $\pi = 0.0062$,
but note that the value of $\pi$ does not affect the number of encounters
per person.

\begin{figure}
\centering
\includegraphics[width = 0.8\linewidth]{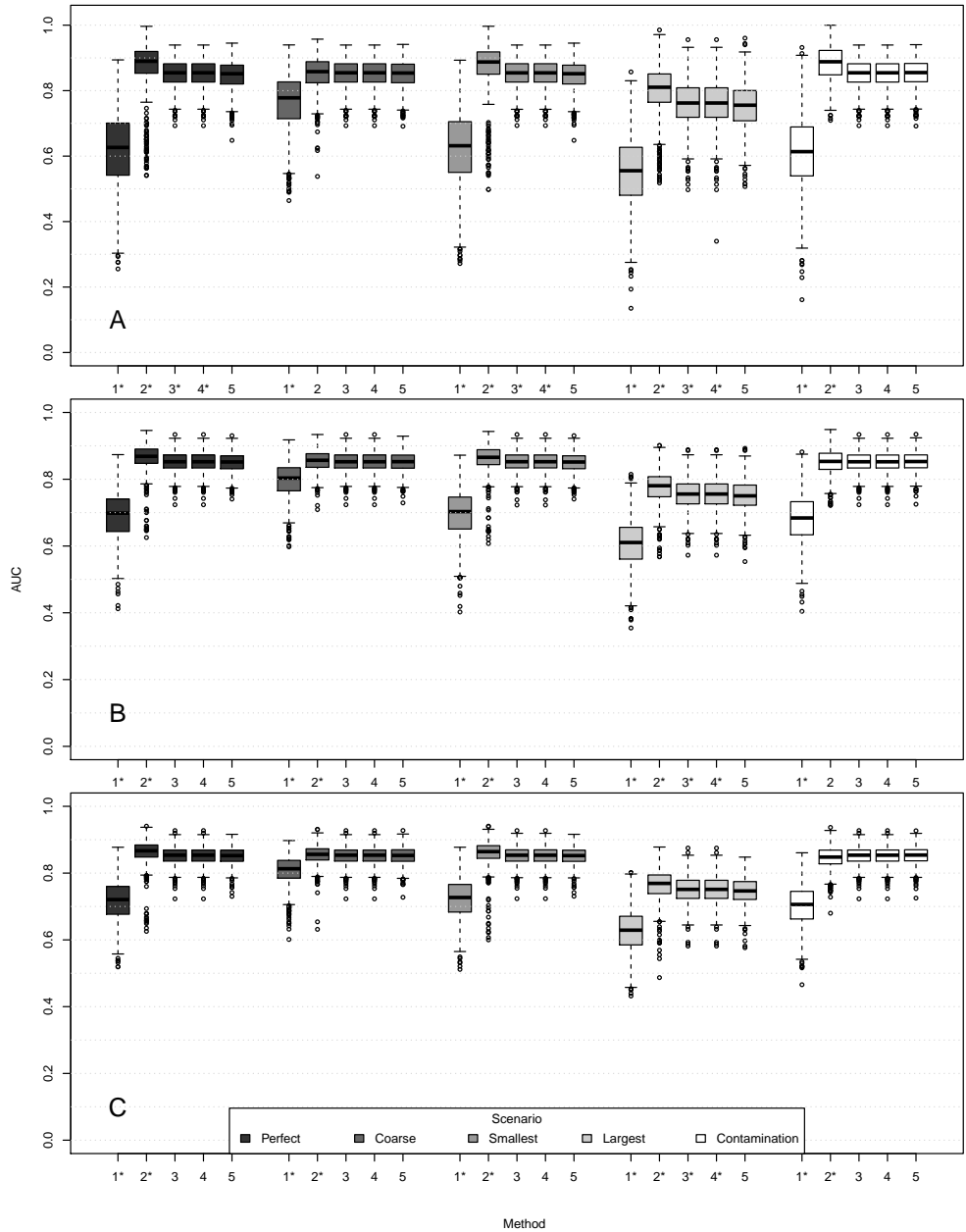}
\caption{\textbf{AUCs from 1,000 replications, single cluster.}
1 = Multiple logistic regression trained on Wave 1 outcome;
2 = Multiple logistic regression trained on Wave 2 outcome;
3 = Simple logistic regression trained on Wave 1 outcome;
4 = Simple logistic regression trained on Wave 2 outcome;
5 = New method.
An asterisk indicates that that method was significantly different
from the new method at the 0.05 level (two-sided).
In panel (A) $\pi = 0.0062$, 
in (B) $\pi = 0.0110$, and 
in (C) $\pi = 0.0143$.}
\label{figure:boxplotaalt}
\end{figure}

AUCs are presented in Figure \ref{figure:boxplotaalt}.
Figure \ref{figure:boxplotaalt}A displays the AUCs for $\pi = 0.0062$,
Figure \ref{figure:boxplotaalt}B displays the AUCs for $\pi = 0.0110$, and
Figure \ref{figure:boxplotaalt}C displays the AUCs for $\pi = 0.0143$.
In general, as $\pi$ increases, the variability of the performance decreases.

For each value of $\pi$ and in each scenario,
multiple logistic regression trained on the Wave 1 outcome performs worst;
simple logistic regression performs about as well as
or slightly better than the new method,
regardless of whether it is trained on the Wave 1 or Wave 2 outcome;
and multiple logistic regression trained on the Wave 2 outcome performs best.

\subsubsection{Discussion}

As $\pi$ decreases, new infections become more rare,
making the relationship between the venues and the outcome
more dependent on chance.
This explains the increasing variability of the risk estimators
with decreasing values of $\pi$.

The similarity of the performance of the new method
with the simple logistic regressions seems to indicate that
an individual's pattern of venue visitation is not as important
as his total number of sexual encounters.
This is corroborated by the performance of the multiple logistic regressions,
which seem to overfit the outcome to the data.
This overfitting causes the multiple logistic regression trained on the
Wave 1 outcome to display the worst performance
and the multiple logistic regression trained on the Wave 2 outcome
to display the best performance.
Another piece of evidence for the primacy of number of sexual encounters
is the Contamination scenario.
Here, the reported pattern of venue visitation differs greatly
from the true pattern of venue visitation,
but each person's total number of encounters remains the same.
The performance of each method is about the same in this scenario
as in the Perfect scenario,
indicating that the total number of encounters is what matters.
The Largest scenario provides a contrast to the Contamination scenario;
here, the three venues with the most sexual encounters are not reported.
This causes a decrease in the total number of sexual encounters
reported by many participants,
and the performance of all five risk estimators suffers.

The distribution of the number of venue visits per person
is approximately the same in the simulation as in the dataset
on which it is based.

\subsection{Simulation Study 2}

The following simulation study was intended to address the
dense clustering of the venues demonstrated in Figure \ref{figure:venue}.
A new dataset was created, this time with two clusters
and a much lower HIV prevalence in one of the clusters.

\subsubsection{Methods}

This simulation study was exactly the same as the first,
with three exceptions.
First, the 579 rows of the dataset were duplicated,
generating a new dataset of 1158 participants.
For the second group of 579 participants,
the ten largest venues were renamed.
This meant there were 40 total venues,
with the first 579 participants only overlapping with the second 579
participants at the twenty smallest venues.
%The resulting venue-to-venue graph is in Figure \ref{figure:venue}B.
Second, this second group of participants was also modified in that
each participant who was HIV-positive at Wave 1
was changed to be HIV-negative at Wave 1 with probability $0.75$.
As a result,
the second group had an HIV prevalence at Wave 1 of $0.0898$,
whereas the first group had an HIV prevalence at Wave 1 of $0.3748$.
Third, the population size $N$ was set to $1158\times 5 = 5790$.
This new dataset formed the basis for the second simulation study
in that in each replication,
1158 participants were drawn uniformly at random from it.

\subsubsection{Results}

\begin{figure}
\centering
\includegraphics[width = 0.8\linewidth]{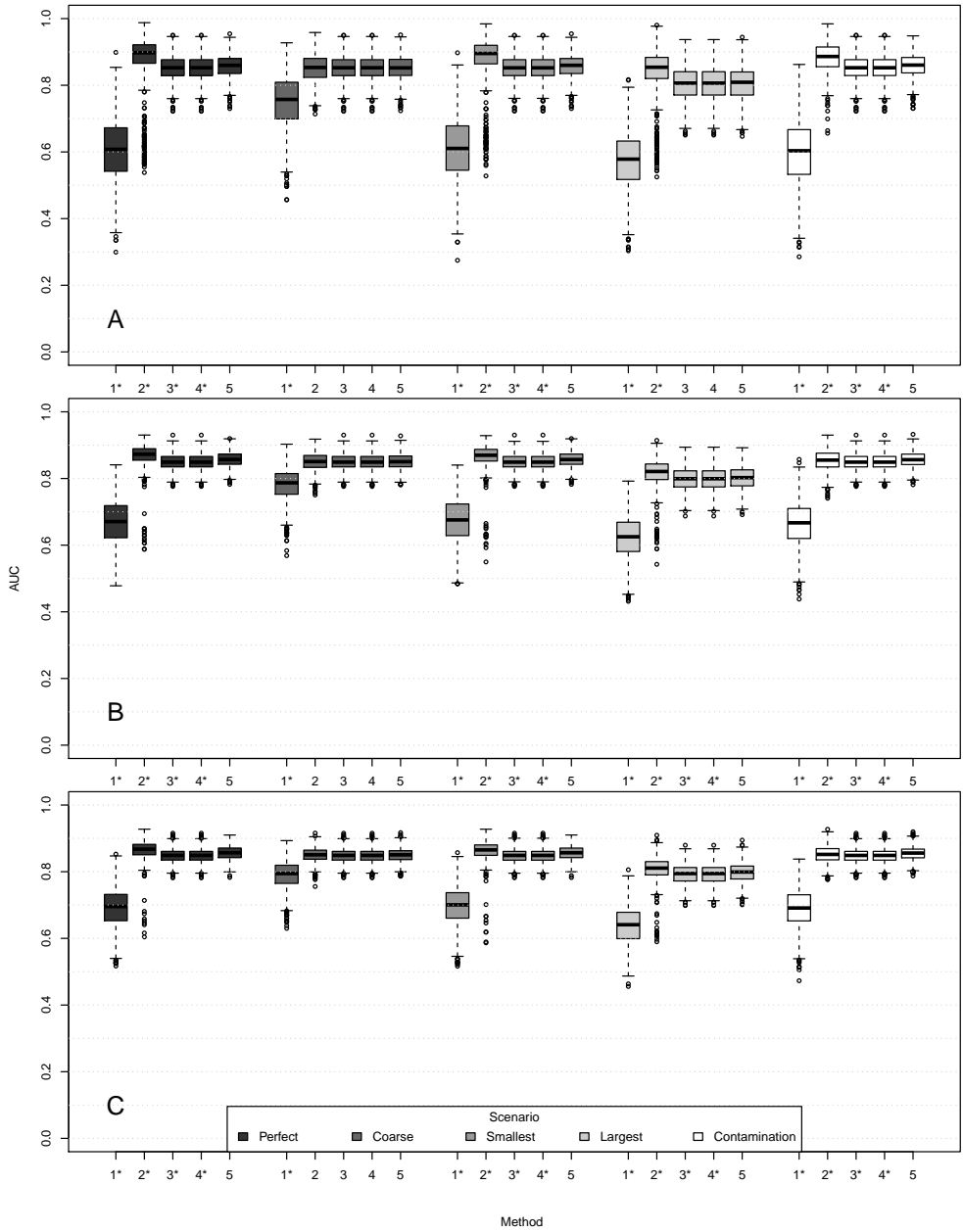}
\caption{\textbf{AUCs from 1,000 replications, two clusters.}
1 = Multiple logistic regression trained on Wave 1 outcome;
2 = Multiple logistic regression trained on Wave 2 outcome;
3 = Simple logistic regression trained on Wave 1 outcome;
4 = Simple logistic regression trained on Wave 2 outcome;
5 = New method. 
An asterisk indicates that that method was significantly different
from the new method at the 0.05 level.
In panel (A) $\pi = 0.0062$, 
in (B) $\pi = 0.0110$, and 
in (C) $\pi = 0.0143$.}
\label{figure:boxplotjalt}
\end{figure}

AUCs are presented in Figure \ref{figure:boxplotjalt}.
Figure \ref{figure:boxplotjalt}A displays the AUCs for $\pi = 0.0062$,
Figure \ref{figure:boxplotjalt}B displays the AUCs for $\pi = 0.0110$, and
Figure \ref{figure:boxplotjalt}C displays the AUCs for $\pi = 0.0143$.
In general, as $\pi$ increases, the variability of the performance decreases.

For each value of $\pi$ and each scenario,
multiple logistic regression trained on the Wave 1 outcome performs worst;
simple logistic regression performs about as well as
or slightly worse than the new method,
regardless of whether it is trained on the Wave 1 or Wave 2 outcome;
and multiple logistic regression trained on the Wave 2 outcome performs best.

\subsubsection{Discussion}

The pattern of results 
differs from that for the previous simulation study
in that the new method performs better than the two
simple logistic regressions.
This seems to indicate that the pattern of venue visitation
is more relevant than the simple total number of sexual encounters per person.
The superiority of the new method over multiple logistic regression
trained on the Wave 1 outcome also indicates that the new method
is not overfitting.

\end{document}